\documentclass[12pt,a4]{article}
\begin{document}

\title{Difference between insulating and conducting states}
\author{E.K.Kudinov\\ \normalsize Ioffe Physico-Technical Institute,
St-Petersbourg, Russia}
\date{ }
\maketitle

\begin{abstract}
It is proposed to used, as a basic property specifying the
differen\-ce between an insulator and a conductor, a static phenomenon,
namely the field effect absent in the former, but present in the latter.
The absence or present of the field effect is closely associated with
the nature of the homogenious linear response to a static electric
field: for an insulator, it is finite; for a conductor, it depends on
the volume $V$ and tends to infinity with $V$. The fluctuation-dissipation
theorem makes it possibile to relate the nature of this response to the
mean square fluctuation $<d^2>$ of the dipole moment. In an insulator,
$\left(<d^2>/V\right)_{V\to\infty}$ is finite; in a conductor, it is
infinite. Thus, in order to ascertain whether a given state insulating
or conducting, it is sufficient to investigate $<d^2>$ at temperatures
close to zero. This in turn amounts to studying  the behavior of the
mean square fluctuation $<\Delta N^2>$ of the number of carriers and
their static pair correlation function. The procedure is illustrated
by a number of examples. In particular, the insulating state is
realized in the Hubbard model for the one-half filling. Some notable
properties of the supercondctivity state have remarked.

\end{abstract}
\vskip 0.5cm
\section{Preface}
This article had published in 1991
(E.K.Kudinov, Fisika Tverdogo Tela {\bf 33}, 2306 (1991); [in English:
Sov.Phys. Solid State {\bf 33}, 1299 (1991)]). It have considered
a distinction between insulating and conducting states which based
on the first principles. In the last time a whole series of the articles,
which considered the same problem (see, for example:
Resta R., Sorella S.  cond-mat/ 9808151.)
without mentoining of the aforesaid article was appeared.
For that reason author belives, that a submission this
manuscript in cond-mat were be resonably.
\section*{}
The band model has been able to represent both
conducting and insulating states of a crystal in terms of band occupations
for the system of the noninterating electrons. As early as 1937, however,
it was noted \cite{boe} that this model does not describe several insulating
crystals. A large number of such materials are now known.
They are compounds of metals with parity filled $d$ or $f$ shells, and
the corresponding bands are only partly occupied \cite{ad,met,god,mot1}.
Since the end of
the 1950s, they have been much studied on account of their particular
features: magnetic and structural transitions, insulator--conductor
transitions, intermediate valence, "heavy" fermion effects, and so on.
In particular, the high-temperature superconductors belong to this class.

Mott \cite{mot2,mot3} has given a qualitive explanation of the
insulating nature of such materials. If the $d$ or $f$ orbital overlap
is small, the electrons should be described by (atomic type) functions
localized at sites; the Coulomb repulsion then causes an effective
attraction of a localized electron and hole, which can form an
electrically neutral bond state and carry no current. The formation of
current exitations is due to the ionisation of such a state (finite
activation energy), which is also responsible for the insulating state
(Mott insulator).The electron interaction evidently has a decisive role
in this pictures, and the question naturally arises of how to formulate
a criterion to differentiate an insulator from a conductor in the
general case, without using a model.

The model proposed in 1964 by Hubbard \cite{hub,her}, which took into
account only one-site Coulomb repulsion, gave rise to an enormous number
of papers, such as Refs. \cite{bri,khom, ber, zai1,zai2}, since it appeared
that the simplicity
of the Hubbard Hamiltonian would allow a description not only of the
insulating and conducting states, but also of the Mott transition
between them.  It was found, however, that the results obtained are not
easly interpreted, mainly for lack of a general
insulator--conductor criterion (for which inadequately justified
assertion have often been substituted).

The criterion must evidently reflect the specific of the ground state of the
system ($T=0$). Those so far proposed fall into two classes: 1) the presence
of a gap in the current exitation spectrum (the band gap in the band theory,
the positive twin--hole pair exitation energy in a Mott insulator), 2)~based
on the Kubo expression for the complex polarisability $\kappa(\omega)$,
a conductor having \cite{mot1,von,kohn} a singularity  of
$\kappa(\omega)$ as $\omega\to 0$.
Class 1 based on the properties of the exitation spectrum and can be
used only with a specific and fairly clear model. The disadvantage of
class 2 is that the necessary properties of the ground state are
determined by the reaction to an external perturbation, so that a
higher-rank problem, the kinetic one, has to be contemplated. The
present paper proposed a criterion based entirely on the static
properties of the ground state.

\section{Qualitative treatment}
Our approachis based on the substancial difference between the linear
responses of an insulator and a conductor (at sufficiently iow temperatures)
to a static homogeneous electric field $\bf E$. For an insulator with a finite
volume $V$, the field inside the body is nonzero and induces a dipole moment
${\bf P}=\kappa_{0}\bf E$ per unit volume, the polarizability
$\kappa_0$ per
unit volume being finite and independent of $V$. In a conductor, there
is a redistribution of charge, and equilibrium corresponds to a
spatially inhomogeneous charge distribution, which reduce to zero the
field acting within the volume (field effect). This inhomogeneity has to
be taken into account from the start when formulating the problem of the
response.

The response can, however, be formally calculated in either case on the
assumption that the final state is a homogeneous one (homogeneous linear
response). For an insulator, this is true, and a reasonable value of
$\kappa_0$ is obtained. The corresponding calculation for a conductor is
bound to show that the problem is incorrectly formulated, by giving an
"anomalous" expression for $\kappa_0$. For an ideal charged Fermi gas,
one easly finds
\begin{equation}
\label{1}
\kappa_{0}=-\frac{e^2}{V}\sum_{\bf kk'}\big|x_{\bf kk'}\big|^2
\frac{n_{\bf k}-n_{\bf k'}}{\varepsilon
_{\bf k}-\varepsilon_{\bf k'}},
\end{equation}
where $x_{\bf kk'}$ are the matrix elements of the coordinate $x$ between
states with the wave vectors $\bf k$ and $\bf k'$ ($x_{\bf kk}=0$,
corresponding to neutrality of the system as a whole), $\varepsilon_{\bf k}=
\hbar^2 k^2/2m$, and $n_{\bf k}$ is the Fermi distribution. A simple
calculation gives, for $T=0$,
\begin{equation}
\label{2}
\kappa_0=\frac{1}{10\pi^2}\left(\frac{3}{4\pi}\right)^{2/3}
\frac{k_c}{r_B}V^{2/3}+\frac{1}{k_{c}r_{B}}O(V^0),
\end{equation}
\noindent where $r_B=\hbar^2/me^2$, $k_c$ is the Fermi momentum, and
$(r_B/k_c)^{1/2}$ is the Thomas--Fermi screening distance. The anomaly
is that $\kappa_0$ depends on the volume: $\kappa_0\to\infty$
as $V\to\infty$; a similar anomaly, as $V^{2/3}$, is found for a
superconductor in the BCS model. The corresponding expression for a band
insulator at $T=0$ gives $\kappa_0$ independent of $V$; there is
no field effect (at $T>0$, an anomalous term occurs as in Eq. (\ref{2}),
proportional to $\exp(-E_g/kT)$, where $E_g$ is the gap width).

Since the polarizability is expressed in terms of the dipole moment
correlation function $<d(t)d> (d\equiv d_x)$ one can expect that the
anomaly will occur also in the corresponding static quantity, the mean
square fluctuation $<d^2>$ of the dipole moment (it is postulated that
there is no ferroelectric ordering, and so $<d>=0$); $<d^2>\sim V$ for
an insulator, but $<d^2>\sim V^{1+\gamma}$ with $\gamma >0$ for a
conductor.

In the limit $V\to\infty$ it is resonable to put all materials in two
classes according to the nature of the homogeneous linear response:
1) $\kappa_0$ is finite (insulator), 2)~$\kappa_0$ is infinite (conductor);
that is, the classification is based on the absence or presence of the
field effect. This is in agreement with the presence of a pole at
$\omega=0$ of the complex polarisability of a conductor \cite{lan1}. In
accordance with the casuality condition, the pole term in
$\kappa(\omega)$ must have the form
$$
{\rm const}\frac{i}{\omega +i\delta}={\rm const}\left(\pi\delta (\omega)+
i\frac{\cal P}{\omega}\right),
$$
where $\cal P$ denotes the principal value; thus, formally, $\kappa_0=
\kappa'(0)=\infty$ (note: $\kappa'(0)$, not
$\kappa'(\omega)\big|_{\omega\to 0}$). We can suppose that class 1 has a
finite value of $\lim_{\,V\to\infty}(<D^2>V^{-1})$ as $T\to 0$, but
class 2 has an infinite one; that is, $d$ has normal fluctuations in
case 1 and anomalous ones in case 2. It will be proved that this
hypothesis follows from the fluctuation-dissipation theorem.

\section{Relationship between the static homoge\-neous response and the
static fluctuations of the dipole moment}
Since the difference between an insulator and a conductor depends on the
specific nature of the ground states, we will everywhere consider
temperatures so low that the contribution from the "gap" modes $e^{-E/kT}$
with $E>0$ is negligible.

The fluctuation-dissipation theorem, in the form
of the Callen--Welton relationship for $\kappa(\omega)$, \cite{lan1} is
\begin{equation}
\label{3}
\frac{\hbar}{\pi}\kappa^{\prime\prime}\coth\frac{\hbar\omega}{2kT}
=\left(\frac{1}{V}\sum_{mn}e^{-E_{n}/kT}|d_{nm}|^2
[\delta(\omega-\omega_{nm})
+\delta(\omega+\omega_{nm})]\right)\Bigg|_{V\to\infty},
\end{equation}
$$
\omega_{nm}=\frac{E_{n}-E_{m}}{\hbar},
$$
where $E_n$ is the energy of steady state number $n$ of the system. (Since
the theorem assumes the energy spectrum of the system to be continuous, we
suppose that the limit $V\to\infty$ has be taken.) Integration of the
right-hand side of Eq. (\ref{3}) over $\omega$ from 0 to $\infty$ gives
$$
\int_{0}^{\infty}d\omega\left(\frac{1}{V}\sum
e^{-E_{n}/kT}|d_{nm}|^2[\delta(\omega-\omega_{nm})
+\delta(\omega+\omega_{nm})]\right)\Bigg|_{V\to\infty}
$$
\begin{equation}
\label{4}
=\left(\frac{1}{V}\sum
e^{-E_{n}/kT}|d_{nm}|^2\right)\Bigg|_{V\to\infty}
=\lim_{V\to\infty}(<d^2>V^{-1}).
\end{equation}
Thus, if $d$ has normal
fluctuations, the integral $\int_0^\infty$ of the left-hand side of Eq.
(\ref{3}) converges; if anomalous fluctuations, then it diverges. This is
valid for infinitesimal $T$ values. The only possible singularity of
$\kappa^{\prime\prime}(\omega)$ is $\omega=0$, and so the convergence
of the integral depends on the behavior of
$\kappa^{\prime\prime}(\omega)$ as $\omega\to0$\footnote{As
$\omega\to\infty,\, \kappa^{\prime\prime}$ must decrease as $\omega^{-2}$,
and there is convergence at the upper limit.}. Let us first take the two
limiting cases.

a) A normal insulator (in which we include an intrisinc semiconductor)
has $\kappa^{\prime\prime}(\omega)$ an analytic function of $\omega$.
The integral
$$
J(T)=\int_0^{\infty}\kappa^{\prime\prime}
\coth\frac{\hbar\omega}{2kT}\,d\omega
$$
is convergent, and as $T\to0$ it tends to a finite limit
$\int_0^{\infty}\kappa^{\prime\prime}\,d\omega$. Hence thoroughout
the temperature range concerned, $d$ has normal fluctuations. The static
polarizability is then finite, since the integral converges in the
Kramers -- Kronig relationships
\begin{equation}
\label{5}
\kappa^{\prime}(0)=\frac{1}{\pi}\int_{-\infty}^{+\infty}
\kappa^{\prime\prime}(\zeta)\frac{\cal P}{\zeta}\,d\zeta.
\end{equation}
b) A normal metal. Here, $\kappa(\omega)$ has a pole term:
\begin{equation}
\label{6}
\kappa(\omega)={\bar\kappa}(\omega)+
\frac{i\sigma_0}{\omega+i\delta},\quad(\delta>0,\,\delta\to0),
\end{equation}
${\bar\kappa}$ has no singularity; $\sigma_0$ is the static
conductivity. (It is assumed that $\sigma_0$ is independend of $T$.)
The integral $J(T)$ diverges for all $T$, including $T=0$. Hence $d$
has anomalous fluctuations. It follows from Eq. (\ref{6}) that
$\kappa^{\prime}(\omega)$ then has a singular term
$\varphi\sigma_0\delta(\omega)$; thus, the homogeneous linear response
is anomalous, $\kappa^{\prime}(0)=\infty$.

A smooth insulator--metal transition can be formally represented by
writing $\kappa^{\prime\prime}(\omega)$ in the form
$\bar\kappa^{\prime\prime}(\omega)+
\kappa_c^{\prime\prime}(\omega)$
\begin{equation}
\label{7}
\kappa_c^{\prime\prime}(\omega)=a\frac{\omega}{|\omega|}
|\omega|^{\alpha},\quad 1\geq\alpha\geq-1
\end{equation}
($a>0$) and $\alpha=1$ for an insulator, $\alpha=-1$ for a metal.
The integral of the lefthand side of Eq. (3) is written as
\begin{equation}
\label{8}
J(T)=\int_0^\infty\kappa^{\prime\prime}
\left(\coth\frac{\hbar\omega}{2kT}-1\right)\,d\omega+\int_0^\infty
\kappa^{\prime\prime}\,d\omega.
\end{equation}
We assume that $a$ and $\alpha$ are independent of $T$. Then, as
$T\to0$,
$$
J=a\int_0^{\infty}\omega^{\prime\prime}
\left(\coth\frac{\hbar\omega}{2kT}-1\right)\,d\omega
+\int_0^{\infty}\kappa^{\prime\prime}\,d\omega
$$
\begin{equation}
\label{9}
=a\left(\frac{2kT}{\hbar}\right)^{1+\alpha}
\int_0^{\infty}x^{\alpha}(\coth x-1)\,dx
+\int_0^{\infty}\kappa^{\prime\prime}\,d\omega.
\end{equation}
\indent1) When $\alpha>0,\quad J(T\to0)\quad$ tends\  to\  the\  finite
\ limit
$\quad\int_0^{\infty}\,\kappa^{\prime\prime}\,d\omega$;\protect\\
$\qquad
\hbox{$(<d^2>/V)_{\infty}$}$, and, by Eq. (\ref{7}),
$\quad\kappa_0=\kappa^{\prime}(0)\quad$ are finite, even for
$\quad T=0$.

\indent2) When $0>\alpha>-1, J$ is infinite for all
nonzero $T$ in the range considered, the fluctuations of $d$ are
anomalous when $T\neq0,\,\kappa_0=\infty$ for all $T$ including
$T=0$, and $\sigma_0=0$ in this range of $\alpha$ values. We thus see
that when $\alpha>0$ the homogeneous linear response is that of an
insulator ($\kappa_0$ finite), and $(<d^2>/V)_\infty$ also is finite.
When $\alpha<0$, the homogeneous linear response corresponds to
$\kappa_0=\infty$ (field effect) and $(<d^2>/V)_{\infty}=\infty$ for
nonzero $T$ (and also for $T=0$ when $\alpha=-1$)\footnote{When
$-1<\alpha<0$, there is an expression $\sim T^{1+\alpha}\cdot\infty$ on
the right of Eq. (\ref{9}), and it has not been possibile to find a correct
passage to $T=0$ so as to determine $(<d^2>/V)_{T=0}$.}. This is the
justification for the hypotesis advanced at the end of Sec. 1, that the
static insulator reaction $\kappa_0$ and the fluctuations of $d$ are
in one-to-one correspondence as regards their nature. The kinetic nature
of $\kappa^{\prime\prime}(\omega)$ acts only as an intermediate link
between these two static characteristics.

This argument justifies the division of substances into two classes
(Sec.~ 1). Classes 1 and 2 correspond to $\alpha>0$ and $-1<\alpha<0$
respectively; the nature of the insulator reaction $\kappa_0$ is in
one-to-one relationship with the behavior of $<d^2>/V$ as $V\to\infty$.
The physical interpretation of class 1 is that the electrons have a
finite motion and so there is no field effect. In class 2, their
motion is infinite, they can go to macroscopic distances, and the
field effect can occur\footnote{When $\alpha>-1,\,\sigma_0=0$, but this
means only that the random walk is not Marcovian.}.
\indent The range of $\alpha$ values between 1 and -1 can apparently
occur only near an insulator--conductor transition; such a transition
by the occurence of a branch point is probably the "smootest" such
transition. The basic criterion (the nature of the static
fluctuations) then retains its meaning even for ideal nonergodic
modeles such as those where the particles do not interact.

\section{Static fluctuations of the dipole moment}
The expression for $<d^2>$ can be put in a clear form.
For simplicity,
let us consider a homogeneous electron gas in a finite volume $V$.
The dipole moment operator $\hat d=\hat d_x$ is
\begin{equation}
\label{10}
\hat d=\int_V x\hat n({\bf r})\,d{\bf r},\quad
\hat n({\bf r})=\sum_{\sigma}\psi_{\sigma}^+({\bf r})
\psi_{\sigma}({\bf r}),
\end{equation}
where $\psi_\sigma^+$ and $\psi_\sigma$ are the Fermi field operators
and $\sigma$ is the spin component.
The coordinates are chosen so that
$\int_V x\,d{\bf r}=0$ (neutrality condition; the origin is taken at the
point where the dipole moment of the positive charges is zero). Then,
\begin{equation}
\label{11}
\frac{1}{V}<d^2>=\frac{1}{V}\int_V<\Delta\hat n({\bf r})
\Delta\hat n({\bf r^\prime})>\,d{\bf r}d{\bf
r^{\prime}},\,
\end{equation}
$$
\Delta\hat n({\bf r})=\hat n({\bf r})-n,\quad
n=<\hat n({\bf r})>.
$$
It is known \cite{lan2} that the density correlation function
${<\Delta\hat n({\bf r})\Delta\hat n({\bf r^{\prime}})>}$ has a
delta-function singularity.\footnote{To make a correct
allowance for this, the field operators must always be put in the normal
order.} This can be separated by writing
\begin{equation}
\label{12}
<\Delta\hat n({\bf r})\Delta\hat n({\bf r^{\prime}})>
=n\delta({\bf r-r^\prime})
+\sum_{\sigma\sigma'}<\psi_{\sigma}^+({\bf r})\psi_{\sigma'}^+({\bf
r^\prime}) \psi_{\sigma'}({\bf r^\prime})\psi_{\sigma}({\bf r})>
-n^2.
\end{equation}
We set
$$
K({\bf rr^\prime})
=<\Delta\hat n({\bf r})\Delta\hat n({\bf r^{\prime}})>
=n\delta({\bf r-r^\prime})
$$
\begin{equation}
\label{13}
+\sum_{\sigma\sigma'}<\psi_{\sigma}^+({\bf r})\psi_{\sigma'}^+({\bf
r^\prime}) \psi_{\sigma'}({\bf r^\prime})
\psi_{\sigma}({\bf r})>.
\end{equation}
This $K({\bf rr^\prime})$ tends to zero as $|\bf r-r^\prime|\to\infty$,
and $K({\bf rr^\prime})=K({\bf r^\prime r})$. Then
\begin{equation}
\label{14}
\frac{1}{V}<\hat d^2>=n\frac{1}{V}\int_V x^2\,d{\bf r}
+\frac{1}{V}\int_V xx^\prime K({\bf r-r^\prime})\,
d{\bf r}d{\bf r^\prime}.
\end{equation}
We substitute here
$xx^\prime=(-1/2)(x-x^\prime)^2+(1/2)(x^2+x^{\prime\,2})$:
$$
\frac{<d^2>}{V}=\frac{n}{V}\int_{V}x^2\,d{\bf r}+\frac{1}{2V}
\int_{V}(x^2+x^{\prime\,2})K({\bf rr^{\prime}})\,d{\bf r}d{\bf
r^{\prime}}
$$
\begin{equation}
\label{15}
 -\frac{1}{2V}\int_{V}(x-x^{\prime})^2 K({\bf
rr^{\prime}})\,d{\bf r}d{\bf r^{\prime}},
\end{equation}
and
$\int_{V}K({\bf rr^{\prime}})\,d{\bf r^{\prime}}$ is
\begin{equation}
\label{16}
\int_{V}K({\bf
rr^{\prime}})\,d{\bf r^{\prime}} =\sum_{\sigma}<\psi^+_{\sigma}({\bf
r})\hat N\psi^{\null}_{\sigma}({\bf r})> -Vn^2.
\end{equation}
Here, $\hat N=\sum_{\sigma}\int_{V}\psi^+_{\sigma}
\psi^{\null}_{\sigma}\,d{\bf r}$ is the total particle number operator
in the volume $V$:
\begin{equation}
\label{17}
\int_{V}K({\bf rr^{\prime}})\,d{\bf r^{\prime}}=<\hat n({\bf r})\hat N>
-n-Vn^2.
\end{equation}
Since the system is homogeneous, $<\hat n({\bf r})\hat n>$ is independent
of $\bf r$, and therefore
\begin{equation}
\label{18}
<\hat n({\bf r})\hat n>=\frac{1}{V}<\hat N^2>.
\end{equation}
The final result is
\begin{equation}
\label{19}
\frac{<d^2>}{V}=\frac{<\Delta\hat N^2>}{V}\int_{V}x^2\,d{\bf r}
-\frac{1}{2V}\int_{V}(x-x^{\prime})^2
K({\bf rr^{\prime}})\,d{\bf r}d{\bf r^{\prime}},
\end{equation}
$$
\Delta\hat N=\hat N-N,\quad N=<\hat N>.
$$
The first term on the right here is of order of $V^{2/3}$, on the
assumption that $\hat N$ has normal fluctuations (that is, far from
any first-order transition point). Its significance is straightforward:
these are dipole moment fluctuations of the charged particle system,
on the assumption that the small macroscopic volumes are statistically
independent; and it follows from elementary arguments. The second term
is  anomalous if the correlation function $K({\bf rr^{\prime}})$ does
not decrease sufficiently rapidly as $|{\bf r-r^{\prime}}|\to\infty$.

There is reason to suppose that always $<\Delta\hat N^2>=0$ in a normal
system at $T=0$; that is, the first term in Eq. (\ref{19}) is zero for $T=0$.
A "normal" system is described by a Gibbs distribution, i.e., by a
density matrix $\rho={\rm const}\cdot\exp[-(\hat H-\mu N)/kT]$. At
$T=0$, the system is in a state $\Phi_0$ corresponding to the lowest
eigenvalue of the operator $\Omega=\hat H-\mu\hat N$ for a given $\mu$.
Since $\hat N$ is an integral of the motion, the number of particles is
a good quantum number: all eigenstates of $\Omega$, including $\Phi_0$,
are  states with a given number of particles. Hence, in each such state,
including $\Phi_0$, the dispersion of the particle number is zero,
$<\Delta\hat N^2>=0$, and a nonzero $<\Delta\hat N^2>$ can occur only by
a spread over different quantum numbers. When $T=0$, however, the system
is in the ground state only, and $<\Delta\hat N^2>_{T=0}=0$.

When there is long-range order, the Gibbs distribution no longer
describes the state of the system; the true distribution function
is found by including infinitesimal terms in the Hamiltonian, and
these break the original symmetry. The fluctuations at $T=0$ are
determined by the specific nature of the broken-symmetry state.

The representation of $<d^2>/V$ in the form (\ref{19}) can be derived
also for a spatially periodic system (more precisly, it is the form of
the terms in $<d^2>$ that are responsible for the presence of anomalous
fluctuations), and for a disordered system under certain plausible
assumptions. It reduces the problem of the $d$ fluctuations to that of
finding $<\Delta\hat N^2>$ and determining the behavior of the
correlation function $K({\bf rr^{\prime}})$.
\section{Examples}
\subsection*{a) Ideal charged Fermi gas}
Here,
$$
<\Delta\hat N^2>=-kTn^2\left(\frac{\partial V}{\partial P}\right).
$$
Since the compressibility is finite, the first term in Eq. (\ref{19}) is
zero for $T=0$. The function $K(\bf rr^{\prime})$ is known \cite{lan2};
its leading term for $T=0$ is
\begin{equation}
\label{20}
K({\bf rr^{\prime}})=K({\bf r-r^{\prime}})
=-\frac{3n}{2\pi^{2}k_c}\frac{\cos^{2}k_c|{\bf r-r^{\prime}}|}
{|{\bf r-r^{\prime}}|}.
\end{equation}
Note that $K<0$. It is seen that, for $T=0$, $d$ has anomalous
fluctuations:
\begin{equation}
\label{21}
\frac{<d^2>}{V}\sim V^{1/3}.
\end{equation}
As $T$ increases, the decrease of $K$ becomes exponential, but the
\linebreak $<\Delta\hat N^2>$ terms begins to be effective, giving a
$V^{2/3}$ anomaly. The reason for the power law (\ref{20}) is the Fermi step
singularity. The repulsive interaction maintains this singularity,
and therefore does not affect the nature of the singularity at $T=0$.
When the temperature is not zero, this singularity is blurred, causing
an exponential decrease of $K({\bf r-r^{\prime}})$.
\subsection*{b) Band insulator}
Here again, $<\Delta\hat N^2>=0$ at $T=0$, in accordance with the
discussion at the end of Sec. 3. At a nonzero temperature,
$<\Delta\hat N^2>\sim\exp(-E_g/kT)$, and so this term is to be
neglected in the temperature range considered. The correlation function
$K({\bf r-r^{\prime}})$ at $T=0$ decreases exponentially, as follows
from the absence of any singularity within the Brillouin zone on account
of the uniform band occupation. Hence, $d$ has normal fluctuations
even at $T=0$.
\subsection*{c) Superconductor} The ordering specific to a
superconductor (ODLRO) needs fluctuations of $\hat N$ in the ground
state\footnote{In a Bogolubov Bose gas at $T=0,\quad <\Delta\hat
N^2>\neq 0$ also.}. This is in contrast to the "normal" systems a) and
b), where the fluctuations of $\hat N$ are thermodynamic, whereasin a
superconductor they are quantum effects and not zero in the ground
state. For the BCS model
\begin{equation}
\label{22}
\frac{<\Delta{\hat N}^2>}{V}\Bigg|_{t=0}=
\frac{4}{V}\sum_{\bf k}u_{\bf k}^2v_{\bf k}^2=
\frac{1}{V}\sum_{\bf k}\frac{\Delta^2}
{(\varepsilon_{\bf k}-\mu)^2+\Delta^2},
\end{equation}
where $\Delta$ is the gap. In accordance with Eq. (\ref{19}), for $T=0$
we have $<\hat d^2>/V\approx V^{2/3}$. It is easy to see that the
correlation function decrease exponentially, since the denominators
such as $[(\varepsilon_{\bf k}-\mu)^2+{\Delta}^2]^{1/2}$ which depend
on $\bf k$ are analytic for real ${\bf k}$ and are nowhere zero.
We accentuate, that the anomalous character of the dipole moment
fluctuations $<d^2>$, which ensures a field effect in a superconductor,
is realized in view of a fluctuation of  $N$ in the ground state of
the superconductor, but no a slow diminshing of the $K(\bf rr')$
as it occurs in a dielectric.
\vskip0.3cm

The straightforward models are thus in agreement with the insulator --
conductor criterion proposed here.

\section{Hubbard model}
The Hubbard model provides a nontrivial example of using the criterion
formulated here in order to distinguish between insulators and
conductors.

The Hubbard Hamiltonian for nondegenerate orbital states is
\begin{equation}
\label{23}
H=\sum_{{\bf mm'}\sigma}J({\bf m-m'})a_{{\bf m}\sigma}^+
a_{{\bf m'}\sigma}^{\null}+U\sigma_{\bf m}\hat n_{\bf m\uparrow}
n_{\bf m\downarrow}\equiv W+H_0,
\end{equation}
with $U>0$. Let us suppose that the number of electrons is equal to
the number $N$ of sites. The site lattice is assumed centrosymmetric.
The form (\ref{23}) of the Hamiltonian presupposes a certain choice of Fermi
field operators, $\psi_{\sigma}({\bf r})=\sum_{\bf m}\varphi_{\bf m}({\bf r})
a_{{\bf m}\sigma}$, where $\{\varphi_{{\bf m}}(\bf r)\}$ is a set of $N$
ortonormalized orbitals, each localized at a lattice site
$\bf m$\footnote{We assume that, as $|\bf r-m|\to\infty,\quad \varphi_{\bf m}
(\bf r)$ falls exponentially.}. Here, $J$ and $U$ are unambiguously defined.
The density operator $\hat n(\bf r)$ is
\begin{equation}
\label{24}
\hat n({\bf r})=\sum_\sigma \psi_{\sigma}^+({\bf r})
\psi_{\sigma}^{}({\bf r})=\sum_{{\bf mm'}\sigma}\varphi_{\bf m}({\bf r})
\varphi_{{\bf m'}}({\bf r})a_{{\bf m}\sigma}^+
a_{{\bf m'}\sigma}^{\null},
\end{equation}
and the density correlation function is
$$
<\Delta\hat n({\bf r})\Delta\hat n({\bf r'})>=
$$
\begin{equation}
\label{25}
\sum\varphi_{{\bf m}_1}({\bf r})\varphi_{{\bf m}_2}({\bf r})
\varphi_{{\bf m}_3}({\bf r'})\varphi_{{\bf m}_4}({\bf r'})
<a_{{\bf m}_1\sigma}^+a_{{\bf m}_2\sigma}^{\null}
a_{{\bf m}_3\sigma'}^+a_{{\bf m}_4\sigma'}^{\null}>-
n({\bf r})n({\bf r'}),
\end{equation}
$$
n({\bf r})=<\hat n({\bf r})>=\sum\varphi_{\bf m}\varphi_{\bf m'}
<a_{{\bf m}\sigma}^+a_{{\bf m'}\sigma}^{\null}>.
$$
We will take into account only the nearest neighbors $J({\bf m-m'})=
J,\, {\bf m-m'=g;\,g}$ is the vector between nearest-neighbor sites.
The overlap is assumed small $(J/U\ll 1)$, and the temperature range
$J^2/U\ll kT\ll U$ is considered. We can then neglected the spin
ordering and the formation of actual twins and holes (gap excitations).
In this range, the state of the system can be represented to order
$(J/U)^2$ by the wave function\footnote{Strictly, one should use the
density matrix $\rho_0$ corresponding to $\varphi_0$, but taking into
account the equal probabilities of the spin configurations.}
\begin{equation}
\label{26}
\Phi=e^{i\hat S}\Phi_0=\left(1+i{\hat S}-\frac{1}{2}\hat
S^2\right),
\end{equation}
where
\begin{equation}
\label{27}
\hat S=-i\sum_{{\bf mm'}\sigma}\frac{J({\bf m-m'})}{U}
(\hat n_{{\bf m}-\sigma}-\hat n_{{\bf m'}-\sigma})
a_{{\bf m}\sigma}^+a_{{\bf m'}\sigma}^{\null},
\end{equation}
$\Phi_0$ is the homopolar state function $(\sum_{\sigma}
\hat n_{{\bf m}\sigma}\Phi_0=\Phi_0)$ with an arbitrary spin
configuration, and $e^{i\hat S}$ is a unitary operator that
eliminates from Eq.(\ref{23}) the term of the first order in $J/U$.
We can find the correlation function (\ref{25}) as far as $(J/U)^2$.
The operator $ a_{\bf m\sigma}^+a_{\bf m^\prime\sigma}$ there is
transformed by means of Eq. (\ref{27}):
\begin{equation}
\label{28}
a_{\bf m\sigma}^+a_{\bf m^\prime\sigma}\to
a_{\bf m\sigma}^+a_{\bf m^\prime\sigma}+
i[\hat S a_{\bf m\sigma}^+a_{\bf m^\prime\sigma}]-
\frac{1}{2}[\hat S[\hat S a_{\bf m\sigma}^+
a_{\bf m^\prime\sigma}]].
\end{equation}
The quantity (\ref{25}) can then be calculated with the transformed operator (\ref{28})
by averaging over $\Phi_0$ and then over all $2^N$ spin configurations.
The result is
$$
<\Delta\hat n({\bf r})\Delta\hat n({\bf r^\prime})>=
\sum_{\bf m}\varphi_{\bf m}^2({\bf r})
\varphi_{\bf m}^2({\bf r^\prime})\sum_{\bf g}\frac{J^2({\bf g})}
{U^2}-
$$
\begin{equation}
\label{29}
\sum_{\bf mm^\prime}\varphi_{\bf m}^2({\bf r})
\varphi_{\bf m^\prime}^2({\bf r^\prime})\sum_{\bf g}
\frac{J^2({\bf g})}{U^2}\delta_{m-m^\prime,\,g}+\dots
\end{equation}
the dots represent terms arising from $n(\bf r) n(\bf r')$, which are even in
$\bf r,\,r'$ and make no contribution to $<d^2>$. Since, in the absence
of ferroelectric ordering, $\varphi_{{\bf m}=0}^2({\bf -r})=
\varphi_{{\bf m}=0}^2({\bf r})$,
$$
<\hat d^2>=e^2\int xx^\prime <\Delta\hat n({\bf
r})\Delta\hat n({\bf r^\prime})> \,d{\bf r}\,d{\bf r^\prime}=
\sum_{\bf m}m_x^2\sum_{\bf g}\frac{J^2}{U^2}-
$$
\begin{equation}
\label{30}
-\sum_{\bf mg}m_x({\bf
m+g})_x\frac{J^2}{U^2}= -\sum_{\bf mg}m_x
g_x\frac{J^2}{U^2}.
\end{equation}
The right-hand side is zero because $\sum_{\bf m}\bf m=0$ (neutrality
condition). Similarly, $<N^2>=0$. The allowance for the temperature
would give hole-twin states in density matrix $\rho_0$, accompanied by
the gap factor $\exp(-E/kT),\,E\sim U$. The state $\Phi$ is thus, to within
$(J/U)^2$, an insulating state according to our view. This becomes clearer if
one considers the structure of the state $\Phi$, Eq. (\ref{26}). The state $\Phi_0$
is homopolar; $\hat S\Phi_0$ contains one twin and one hole, $\hat S^2\Phi_0$
not more than two of each.  However, it is phisically obvious that in the
ground state of a conductor the number of polar excitations, independend
of $N$, as in $\Phi$, does not make the state conducting. Hence, in any
order of
\noindent
perturbation theory of the type (\ref{26}), the state remains insulating
\footnote{In the calculations, we neglected in Eq. (\ref{30}) the surface contribution,
assuming that for all the sites over which the summation is taken there is
the same number $z=\sum_{\bf g}$ of nearest neighbors. In contrast to this,
the fluctuations in a Fermi gas, proportional to $V^{1/3}$, are volume
fluctuations; according to the meaning of the fluctuation-dissipation theorem,
which involves the limit $V\to\infty$, these fluctuation are the
important ones.}.

The insulating state may therefore contain an admixture of polar states,
expressed by a nonzero
$<\hat n_{{\bf m}\uparrow}\hat n_{{\bf m}\downarrow}>$; in our
approximation, $<\hat n_{\uparrow}\hat n_{\downarrow}>=
(z/2)(J/U)^2$. Attempts to link the transition to the conducting state
with the loss of homopolarity have been made in \cite{ber}, for example.
However, it is evident from the above that the admixture of polar
states does not at all imply the presence of the charges capable
of infinite motion.
\section{Discussion of results}
The proposed criterion to differentiate  an insulator from a conductor depends
the behavior of  the mean square dipole moment fluctuation, a purely static
quantity.   For an insulator, the dipole moment is additive, since its mean
square is \cite{lan2} proportional to $V$. This may be regarded as a
manifestation of
electron localisation relative to the ions at lattice sites.  Anomalious behavior,
$<d^2 >\propto  V^{1+\gamma}$ with $\gamma>0$, corresponds to
delocalisation of electrons. That is, the particular behavior of \linebreak
$(<d^2>/V)_{V\to\infty}$ gives a precise meaning to the intuitive idea of
localized  or delocalized electron states (finite or infinite electron motion).
The nature of the delocalisation is substantially different for a normal
conductor, the anomaly of $<d^2>_{T=0}$ is due to the slow decrease of the correlation function at large distances; in a superconductor, to the presence
of density fluctuations at $T=0$. This will be more fully discussed elsewhere.

 The fluctuation-dissipation theorem makes possible a definite association
between the localization or delocalization and other static quantity, the
reaction $\kappa_0$ to a homogeneous static electric field (homogeneous
linear response), which is fundamentally different for conductors and
insulators\footnote{An attempt to use the Callen-Welton relationship
for the conductivity $\sigma(\omega)$ would not lead to the classification
sought, since the current correlation function $<J^2>$, unlike $<d^2>$,
is always "normal". }.  The kinetic quantity $\kappa''(\omega)$ does not
appear in the final result. It should be emphased that the conducting state is
characterized {\em e contrario}: the quantities $\kappa_0$ and
$<d^2>/V,\, V\to\infty$ are finite for an insulator, are infinite for a
conductor, and so have no meaning.

The transition from an insulator to a normal conductor amounts to a change
in the asymptotic form of $K(\bf rr')$ as $|{\bf r-r'|}\to\infty\,(T=0)$.
It is not accompanied by "strong" fluctuations, and therefore cannot be represented by any order parameter, even conventionally as in a gas-liquid transition. There is, however, a parallel with transitions at $T\neq0$ in
two-dimentional systems \cite{ber1}.

We have in the foregoing extended our treatment to ideal non-ergodic
systems, on the assumption that, if the "kinetic" terms in the Hamiltonian,
responsible for ergodicity, are small, then neglecting them does not affect
on the classification of the system as insulator or conductor.  For a
conductor, the omission of these terms simply emphasizes the singularities
( for example, $\sigma_0$ becomes infinite); for an insulator, it does not
cause a field effect. However, including them is important near an insulator-
conductor transition and also in low-dimensional disorder systems, where
an exponential decrease of $K(\bf rr')$  can occur even with infinitesimal
disorder.

The problem of classifying insulators and conductors is thus reduced to an
investigation of static properties: the mean square fluctuation
$<\Delta N^2>$
of the number of electrons, and the correlation function $K(\bf rr')$ near
$T=0$. This allows also a clear formulation of the insulator-normal
conductor transition problem.

\newpage
\sloppy


\begin{thebibliography}{99}
\bibitem{boe} J. H. de Boer and E. J. W.
Verwey, {\sl Proc.  Phys. Soc.  London} {\bf 49}, 59 (1937).
\bibitem{ad} D. Adler, {\sl Rev. Mod. Phys.} {\bf 40}, 714 (1968).
\bibitem{met} S. Methfessel and D. C. Mattis, in:
{\sl Handbuch der Physik}, Vol. 18,
Part 1, Springer Verlag, Berlin (1968), p. 389.
\bibitem{god} J. B. Goodenough,{\sl Magnetism and the Chemical Bond},
Interscience, New York (1963).
\bibitem{mot1} N. F. Mott, {\sl Metal -- Insulator Transitions}, Taylor and
Francis, London (1974).
\bibitem{mot2} N. F. Mott, {\sl Proc. Phys. Soc. London
Sect. A}62, 416 (1949).
\bibitem{mot3} N. F. Mott, {\sl Rev. Mod. Phys.} {\bf
40}, 677 (1968).
\bibitem{hub} J. Hubbard, {\sl Proc.R. Soc. London Ser. A}{\bf
276}, 238 (1963) (I); {\bf 277}, 237 (1964) (II); {\bf 281}, 401 (1964)
(III).
\bibitem{her} C. Herring, in: {\sl Magnetism}, G. T. Rado and H. Suhl
(eds.), Vol. IV, Academic Press, New York\, (1966).
\bibitem{bri} W. F. Brinkman and T. M. Rice, {\sl Phys. Rev., B}\,{\bf 2},
4302\,(1970).
\bibitem{khom} D. I. Khomskii, {\sl Fiz. Met. Metalloved.}\,{\bf 29},
31\,(1970).
\bibitem{ber} J. Bernasconi, {\sl Phys. Kondens. Mater.}\,{\bf 14},
225\,(1972).
\bibitem{zai1} R. O. Zaitsev, {\sl Zh. Eksp. Teor. Fiz.}\,{\bf 75},
2362\,(1978)[{\sl Sov. Phys. JETP}\,{48}, 1193\,(1978)].
\bibitem{zai2} R. O. Zaitsev, E. V. Kuz'min, and S. G. Ovchinnikov,
{\sl Usp. Fiz. Nauk}\,{\bf 148},603\,(1986), [{\sl Sov. Phys. Usp.}
\,{\bf 29}, 322\,(1986)].
\bibitem{von} S. V. Vonsovskii and M.I.Katznel'son, in: {\sl Problems
of Modern Physics} [in Russian], Nauka, Leningrad\,(1980), p. 233.
\bibitem{kohn} W. Kohn, {\sl Phys. Rev.}\,{\bf 133}, A171\, (1964).
\bibitem{lan1} L. D. Landau and E. M. Lifshitz, {\sl Electrodynamics
of Continuous Media}, Pergamon Press, Oxford\,(1960).
\bibitem{lan2} L. D. Landau and E. M. Lifshitz, {\sl Statistical
Physics}, 2nd ed., Pergamon Press, Oxford\,(1969).
\bibitem{ber1} V. L. Berezinskii, {\sl Zh. Eksp. Teor. Fiz.}\,{\bf 59},
907\,(1970); {\bf 61}, 114\,(1971)[{\sl Sov. Phys. JETP}\,{\bf 32},
493\,(1971);{\bf 34}, 610\,(1972)].
\end{thebibliography}
\end{document}